\documentstyle[12pt]{article}

\title{
{\vspace{-3cm} \normalsize
\hfill \parbox{30mm}{DESY 93-184}   }\\[25mm]
Minimal U(1) gauge fields in two dimensions }

\author{ I. Montvay  \\[4mm]
Deutsches Elektronen-Synchrotron DESY, \\
Notkestr.\,85, D-22603 Hamburg, FRG }

\date{}

\newcommand{\be}{\begin{equation}}
\newcommand{\ee}{\end{equation}}
\newcommand{\bd}{\begin{displaymath}}
\newcommand{\ed}{\end{displaymath}}
\newcommand{\half}{\frac{1}{2}}

\newcommand{\cpt}{\mbox{\rm cpt}}

\begin{document}
\maketitle

\begin{abstract} \normalsize
 Gribov copies of the vacuum in two dimensional compact U(1)
 lattice gauge models are constructed.
 On the basis of this a gauge fixing algorithm is developped,
 wich finds the minimum of the sum of link field squares.
 Numerical experience in a two dimensional Higgs model with fixed
 length scalar field is reported and the extension to three and four
 dimensional U(1) and four dimensional SU(2) gauge theories is briefly
 discussed.
\end{abstract}


\section{Introduction}                                    \label{s1}
 Lattice gauge theory \cite{WILSON} is explicitly gauge invariant
 but gauge fixing is often desirable, for instance, to establish
 relation with perturbation theory.
 Therefore, the properties of the space of gauge orbits has been
 extensively studied both in continuum and lattice formulations
 \cite{GRIBOV,ZWANZI,VANBAA} (for further references on the extensive
 literature see the review \cite{SALMHO}).

 On the lattice, gauge fixing to the Landau gauge can be formulated
 as a minimization problem: one has to find the minimum of the sum of
 link gauge field squares along gauge orbits.
 In most cases the absolute minimum is unique, but the appearence of
 a large number of secondary local minima leads to {\em Gribov
 copies} \cite{GRIBOV}, which are difficult to sort out even is the
 simple case of compact U(1) gauge fields.
 (For recent work on gauge fixing in U(1) and further references see
 \cite{HULSEB,BOMIMUPA,POYEZU}.)

 In the present paper the problem of minimizing the gauge field
 by gauge transformation will be considered in a two dimensional Higgs
 model.
 The lattice action depending on the compact U(1) gauge field
 $U_{x\mu}=\exp(iA_\mu(x)),\; (\mu=1,2)$ and, for simplicity, fixed
 length Higgs scalar field $\phi(x),\; |\phi(x)|=1$ has two parameters:
 the inverse gauge coupling squared $\beta=1/g^2$  and the hopping
 parameter of the scalar field $\kappa$:
\be \label{eq01}
S = \beta \sum_x \sum_{\mu=1, \nu=2}
[1 - cos(F_{\mu\nu}(x))]
-2\kappa \sum_x \sum_{\mu=1}^2 \phi^*(x+\hat{\mu})U_{x\mu}\phi(x) \ .
\ee
 Here the lattice gauge field strength is defined for $\mu,\nu=1,2$ as
\be \label{eq02}
F_{\mu\nu}(x) =
A_\nu(x+\hat{\mu})-A_\nu(x)-A_\mu(x+\hat{\nu})+A_\mu(x) \ .
\ee
 $\kappa=0$ gives pure U(1) gauge theory without interaction with
 the scalar field.
 Real angular variables $-\pi < \theta_{x\mu} \leq \pi$ on the links
 $(x,x+\hat{\mu})$ can be introduced by
\be \label{eq03}
U_{x\mu} \equiv \exp(i\theta_{x\mu}) \ , \hspace{2em}
\theta_{x\mu} \equiv \pi t_{x\mu} = \cpt (A_\mu(x)) \equiv
A_\mu(x)-2\pi \cdot \mbox{\rm NINT}(A_\mu(x)/2\pi) \ ,
\ee
 where NINT() denotes nearest integer and the normalized angle
 $-1 < t_{x\mu} \leq 1$ as well as the function $\cpt()$
 are introduced for later convenience.

 The Lorentz gauge condition $\partial_\mu A_\mu(x) = 0$
 can be latticezed \cite{HULSEB,BOMIMUPA,POYEZU} by looking for extrema
 of
\be \label{eq04}
f_1[U] \equiv \sum_x \sum_{\mu=1}^2 [1 - \cos\theta_{x\mu}]  \ .
\ee
 Another choice, instead of $f_1$, is simply
\be \label{eq05}
f_2[U] \equiv \sum_x \sum_{\mu=1}^2 \theta_{x\mu}^2  \ .
\ee
 A gauge transformation with parameters $\alpha_x$ acts on the link
 angles according to
\be \label{eq06}
\theta^{(\alpha)}_{x\mu} = \cpt\left( \theta_{x\mu} +
\alpha_x - \alpha_{x+\hat{\mu}} \right) \ .
\ee
 Therefore the extremum condition for $f_2$ under gauge transformations
 is
\be \label{eq07}
0 = \sum_{\mu=1}^2 (\theta_{x\mu}-\theta_{x-\hat{\mu},\mu})
= \Delta^b_\mu \theta_{x\mu} \ ,
\ee
 with the backward lattice derivative $\Delta^b_\mu$.
 This is a simple discretization of the continuum Lorentz condition
 $\partial_\mu A_\mu(x) = 0$.
 For small fields in lattice units  $|\theta_{x\mu}| \ll 1$,
 which dominate in perturbation theory, the minimization of
 $f_1[U]$ and $f_2[U]$ is equivalent.
 For large fields the choice is free, but $f_2$ is technically simpler,
 therefore in the present paper $f_2$ will be used.


\section{Gribov copies of the vacuum}                     \label{s2}
 In what follows we always assume periodic boundary conditions on
 the $L^2$ lattice.
 The Gribov copies of the vacuum satisfy, besides the extremum
 condition (\ref{eq07}), the requirement that every plaquette
 and Polyakov line winding around the torus gives a group element 1
 in U(1).
 That is:
\be \label{eq08}
\theta_{x1} + \theta_{x+\hat{1},2} -
\theta_{x2} - \theta_{x+\hat{2},1} = 2\pi n_{x12} \ ,
\ee
and for $\mu=1,2$
\be \label{eq09}
\sum_{k=0}^{L-1} \theta_{x+k\hat{\mu},\mu} = 2\pi l_{x\mu} \ .
\ee
 Here $n_{x\mu\nu}$ and $l_{x\mu}$ are integers.
 Due to $|\theta_{x\mu}| \leq \pi$ the allowed values are
 $-2 \leq n_{x\mu\nu} \leq 2$ and $-L/2 \leq l_{x\mu} \leq L/2$.
 In the real (identically zero: $\theta_{x\mu} \equiv 0$) vacuum
 we obviously have $n_{x12}=l_{x1}=l_{x2}=0$.
 Solutions with some non-zero $n_{x\mu\nu}$ and $l_{x\mu}$ give
 Gribov copies.

 Plaquettes with non-zero values of $n_{x12}$ play an important r\^ole
 in the dynamics of compact U(1) lattice gauge fields
 \cite{DEGTOU,PROCS}.
 Such plaquettes are said to carry a {\em Dirac string}, or a
 {\em gauge field kink}.
 In the present paper we shall call a plaquette with $n_{x12}=1$ a
 {\em kink}, with $n_{x12}=-1$ an {\em antikink}, with $n_{x12}=2$ a
 {\em double kink}, and with $n_{x12}=-2$ a {\em double antikink}.

 Let us define in the general case, for a gauge configuration which
 is not necessarily gauge equivalent to the vacuum, the plaquette
 variables
\bd
\Theta_{x\mu\nu} \equiv \theta_{x\mu} + \theta_{x+\hat{\mu},\nu} -
                        \theta_{x\nu} - \theta_{x+\hat{\nu},\mu} \ ,
\ed
\be \label{eq10}
\theta_{x\mu\nu} \equiv \cpt \Theta_{x\mu\nu}
\equiv \pi t_{x\mu\nu} \ , \hspace{2em}
n_{x\mu\nu} \equiv \frac{1}{2\pi} \left(
\Theta_{x\mu\nu} - \theta_{x\mu\nu} \right) \ .
\ee
 Then the integer valued topological charge $Q$ of the configuration
 (see ref. \cite{GKSW} and references therein) is given by
\be \label{eq11}
Q = \frac{1}{2\pi} \sum_x \theta_{x12}
= \half \sum_x t_{x12} = - \sum_x n_{x12} \ .
\ee
 Since in the vacuum $Q=\theta_{x12}=0$, the sum of the integers
 $n_{x12}$ on the right hand side of (\ref{eq08}) has to vanish for
 every vacuum gauge configuration.

 Note that $n_{x12}$ is not gauge invariant.
 Namely, a gauge transformation where the argument of $\cpt()$ in eq.
 (\ref{eq06}) on some link is greater in absolute value than $\pi$ adds
 a kink-antikink pair to the two plaqettes containing the link.
 At the same time the value of the Polyakov line going through the
 link jumps by $\pm 2\pi$, therefore the integer $l_{x\mu}$ on the
 right hand side of (\ref{eq09}) jumps by $\pm 1$.
 Let us call such a link {\em overtransformed}.
 Since for independent gauge transformations one can restrict the
 parameters to $-\pi < \alpha_x \leq \pi$, starting from the identically
 zero vacuum configuration, one can overtransform every link at most
 once.
 Therefore, a set of overtransformed links with signs uniquely
 determines the values of the integers $(n_{x12},\; l_{x1},\; l_{x2})$
 in eqs. (\ref{eq08}), (\ref{eq09}).
 If the linear system of equations (\ref{eq07}), (\ref{eq08}),
 (\ref{eq09}) has a solution then we obtain a Gribov copy of the
 vacuum, provided that the solution is an admissible link angle
 configuration ($-1 < t_{x1},t_{x2} \leq 1$) and the extremum of $f_2$
 is a minimum.

 The stability of the extrema can be easily seen, because
 the second derivative matrix $D_{yx}$ defined by
\be \label{eq12}
D_{yx} \equiv \frac{\partial^2 f_2[U^{(\alpha)}]}
{\partial\alpha_y \partial\alpha_x}
\ee
 is equal to
\be \label{eq13}
D_{yx} = 4 \delta_{yx} - \sum_{\mu=1}^2
[ \delta_{y,x+\hat{\mu}} + \delta_{y,x-\hat{\mu}} ] \ .
\ee
 As it can be easily proven, this matrix has a single zero
 eigenvalue corresponding to $\alpha_x = {\rm const.}$, which is
 a global gauge transformation.
 All the other eigenvalues are positive, therefore the vacuum
 configurations satisfying (\ref{eq07}) are all stable minima.

 Concerning $|t_{x1},t_{x2}| \leq 1$ at the candidate Gribov copies,
 to obtain a general statement does not seem easy.
 Numerical experience shows that in most cases with $|n_{x12}|=1$
 the values of the normalized link angles $t_{x1},t_{x2}$ are in the
 allowed range.
 As an example, a kink at $(x_1,x_2)=(1,3)$ and antikink at $(5,3)$
 on $8^2$ lattice looks like (\ref{eq14}), (\ref{eq15}) where,
 respectively, the normalized link angles $t_{\mu=1}$ and
 $t_{\mu=2}$ are given.
 The sum of link angle squares in this configuration is
 $f_2=(1011/238)\pi^2$.
 Similar results on larger lattices are too voluminous to be published
 here in detail.
 Just as an example, a kink-antikink pair at $(8,16),(24,16)$ on $32^2$
 gives $f_2=(21.7141\ldots/3.60523\ldots)\pi^2$.

\be \label{eq14}
t_1 =
\left [\begin {array}{cccccccc} {\frac {23}{1904}}&{\frac {79}{1904}}&
{\frac {159}{1904}}&{\frac {215}{1904}}&-{\frac {215}{1904}}&
-{\frac {159}{1904}}&-{\frac {79}{1904}}&-{\frac {23}{1904}}
\\\noalign{\medskip}
{\frac {9}{476}}&{\frac {67}{952}}&{\frac {171}{952}}&
{\frac {229}{476}}&-{\frac {229}{476}}&-{\frac {171}{952}}&
-{\frac {67}{952}}&-{\frac {9}{476}}
\\\noalign{\medskip}
{\frac {23}{1904}}&{\frac {79}{1904}}&{\frac {159}{1904}}&
{\frac {215}{1904}}&-{\frac {215}{1904}}&-{\frac {159}{1904}}&
-{\frac {79}{1904}}&-{\frac {23}{1904}}
\\\noalign{\medskip}
0&0&0&0&0&0&0&0\\\noalign{\medskip}
-{\frac {23}{1904}}&-{\frac {79}{1904}}&-{\frac {159}{1904}}&
-{\frac {215}{1904}}&{\frac {215}{1904}}&{\frac {159}{1904}}&
{\frac {79}{1904}}&{\frac {23}{1904}}
\\\noalign{\medskip}
-{\frac {9}{476}}&-{\frac {67}{952}}&-{\frac {171}{952}}&
-{\frac {229}{476}}&{\frac {229}{476}}&{\frac {171}{952}}&
{\frac {67}{952}}&{\frac {9}{476}}
\\\noalign{\medskip}
-{\frac {23}{1904}}&-{\frac {79}{1904}}&-{\frac {159}{1904}}&
-{\frac {215}{1904}}&{\frac {215}{1904}}&{\frac {159}{1904}}&
{\frac {79}{1904}}&{\frac {23}{1904}}
\\\noalign{\medskip}
0&0&0&0&0&0&0&0\end {array}\right ]_{{{x_1=0\ldots 7},{x_2=0\ldots 7}}}
\ee
\be \label{eq15}
t_2 =
\left [\begin {array}{cccccccc} {\frac {19}{272}}&{\frac {27}{952}}&
-{\frac {15}{272}}&-{\frac {20}{119}}&-{\frac {15}{272}}&
{\frac {27}{952}}&{\frac {19}{272}}&{\frac {39}{476}}
\\\noalign{\medskip}
{\frac {27}{272}}&{\frac {67}{952}}&-{\frac {7}{272}}&
-{\frac {375}{952}}&-{\frac{7}{272}}&{\frac {67}{952}}&
{\frac {27}{272}}&{\frac {101}{952}}
\\\noalign{\medskip}
{\frac {41}{272}}&{\frac {171}{952}}&{\frac {75}{272}}&
{\frac {613}{952}}&{\frac {75}{272}}&{\frac {171}{952}}&
{\frac {41}{272}}&{\frac {137}{952}}
\\\noalign{\medskip}
{\frac {49}{272}}&{\frac {211}{952}}&{\frac {83}{272}}&
{\frac {199}{476}}&{\frac {83}{272}}&{\frac {211}{952}}&
{\frac {49}{272}}&{\frac {20}{119}}
\\\noalign{\medskip}
{\frac {49}{272}}&{\frac {211}{952}}&{\frac {83}{272}}&
{\frac {199}{476}}&{\frac {83}{272}}&{\frac {211}{952}}&
{\frac {49}{272}}&{\frac {20}{119}}
\\\noalign{\medskip}
{\frac {41}{272}}&{\frac {171}{952}}&{\frac {75}{272}}&
{\frac {613}{952}}&{\frac {75}{272}}&{\frac {171}{952}}&
{\frac {41}{272}}&{\frac {137}{952}}
\\\noalign{\medskip}
{\frac {27}{272}}&{\frac {67}{952}}&-{\frac {7}{272}}&
-{\frac {375}{952}}&-{\frac {7}{272}}&{\frac {67}{952}}&
{\frac {27}{272}}&{\frac {101}{952}}
\\\noalign{\medskip}
{\frac {19}{272}}&{\frac {27}{952}}&-{\frac {15}{272}}&
-{\frac {20}{119}}&-{\frac {15}{272}}&
{\frac {27}{952}}&{\frac {19}{272}}&{\frac {39}{476}}
\end {array}\right ]_{{{x_1=0\ldots 7},{x_2=0\ldots 7}}}
\ee

 An exceptional situation is, when the kink and antikink are touching
 each other in a common overtransformed link.
 In this case the solution of the linear equations give for the
 overtransformed link a value outside the allowed range.
 Therefore, the minimum is on the boundary: one has to set the
 overtransformed link to $t=\pm 1$ and solve the equations without
 the ones in (\ref{eq07}) for $x$ at the two endpoints of this link.
 It turns out that the remaining equations have a unique solution
 within the allowed range and on an $8^2$ lattice one obtains
 $f_2=(128/63)\pi^2$.

 In case of a plaquette with double kink the only possibility is
 to have $t_{x1}=t_{x+\hat{1},2}=-t_{x2}=-t_{x+\hat{2},1}=\pm 1$.
 Therefore, these values have to be fixed before solving the linear
 system of equations.
 Correspondingly, one has to omit the equations in (\ref{eq07})
 for points on plaquettes with double kinks.
 The remaining system of equations usually does not have a unique
 solution, therefore one has to look for the minimum of $f_2$ in
 the remaining few variables.
 For instance, for a double kink pair at the same positions as in
 (\ref{eq14}), (\ref{eq15}) one has to fix a single remaining
 variable.
 The solution gives $f_2=(1328940/76409)\pi^2$, a value
 substantially higher than for a normal kink pair.
 This is one of the reasons why in numerical simulations Gribov copies
 with double kinks usually are not important.

\section{Gauge fixing algorithm}                          \label{s3}
 Introducing a small gauge coupling will not qualitatively change the
 Gribov copies.
 Therefore one can characterize them by the kink configuration, as
 in the case of the vacuum.
 The system of linear equations in (\ref{eq07}), (\ref{eq08}),
 (\ref{eq09}) changes, because on the right hand side of (\ref{eq08})
 also the non-integer part $\theta_{x12}$ of the total plaquette
 angle $\Theta_{x12}$ appears.
 Similarly, in (\ref{eq09}) also the non-integer part of the Polyakov
 lines have to be included.

 On smallish lattices one can then determine the values of $f_2$
 at every local minimum and choose the one with the lowest value.
 However, the number of local minima is increasing exponentially
 with the number of lattice points, therefore this is not
 practicable for larger lattices.
 A possibility is to visit a large number of local minima in a
 random walk, and keep always the lowest one.
 Such an algorithm can consist of the following steps:
\begin{enumerate}
\item {\em Gauge cooling;}
\item {\em Removing kink pairs;}
\item {\em Changing kink configuration.}
\end{enumerate}

 In the first step one can apply an iterative algorithm to take the
 gauge configuration to a local minimum.
 One can sweep systematically on sites and minimize $f_2$ locally, or
 choose links randomly and try random gauge transformations at both
 ends and accept the change if $f_2$ gets smaller.

 In the second step one looks for the kink distribution on the
 configuration and tries to remove all kink-antikink pairs.
 A possibility is to choose the closest pairs and perform the opposite
 of the gauge transformation which would create this pair from the
 zero vacuum (see (\ref{eq14}), (\ref{eq15})).
 In praxis, however, it is simpler to work with idealized kinks, since
 the kinks on finite lattices and at finite distances from each
 other are somewhat influencing each other.
 An idealized kink is shown in fig. 1.
 An idealized kink-antikink pair together with the gauge transformation
 creating it from the vacuum is given in fig. 2.
 One can see that on the horizontal and vertical straight sections
 between the kink pair there is a string of overtransformed links,
 which connects the kink and antikink.
 After removing all kink pairs one has still to transform the Polyakov
 loops in such a way that the integer numbers $l_{x\mu}$ on the
 right hand side of eq. (\ref{eq09}) on average vanish in both
 directions ($x_1$ and $x_2$ direction).
 This can be achieved by introducing an appropriate number of strings
 of overtransformed links which are closed by the periodic boundary
 conditions.
 After performing this transformations, on a typical non-vacuum
 configuration at sufficiently small gauge coupling, the configuration
 comes close to a local minimum.
 At zero gauge coupling and for topological charge zero one would create
 in such a way the identically zero vacuum configuration.
 For configurations with non-zero topological charge the result is
 a smooth configuration with ''instantons''.
 Of course, applying step 1. again one can take the configuration
 at any time arbitrarily close to the nearby local minimum.

 The purpose of step 2. is to create a good starting configuration for
 the final step, which is a random walk on local minima in order to
 find with high probability the absolute minimum.
 This is necessary, because on non-vacuum configurations the
 different kink positions give different $f_2$ values and one has to
 find the optimal positions.
 The moving of kinks and antikinks is done by gauge transformations
 corresponding to the desired moves in a zero vacuum background
 (see the example in fig. 2).
 It is also possible that a deeper minimum is achieved by adding again
 a few kink-antikink pairs.
 This can be taken into account during the random walk by allowing for
 kink-antikink pair creation with some small probability.
 Annihilation of kink-antikink pairs occurs if during the random
 moves a kink hits an antikink or vice versa.

 The strategy of the random walk can be chosen differently.
 A good way is to look for a deeper minimum in some given finite number
 of steps.
 If it is found than move the configuration to the new minimum,
 otherwise start again another sequence of moves from the known deepest
 minimum.
 Another possibility is to do a sequence of single steps and to
 allow with some small and decreasing probability also increasing values
 of $f_2$ at the visited local minima (''annealing'').

 I collected some numerical experience with this algorithm in the
 two dimensional Higgs model defined by the lattice action (\ref{eq01})
 on $64^2$ lattice at $\beta=2$ and $\beta=8$ for
 $0 \leq \kappa \leq 0.8$.
 At $\beta=8$ where the plaquette expectation value is about
 $\simeq 0.94$, on most configurations taken from a
 Metropolis-overrelaxation updating the absolute minimum could be found
 in many different subsequent random searches.
 A representative sample of $f_2$ values at the local minima with
 different numbers of additional kink-antikink pairs are contained
 in table 1.
\begin{table}[tb]
\caption{  \label{tb1}
 Comparison of deepest minima in sectors with different numbers
 of kink-antikink pairs $N_p$.
 The value of $f_2$ is given.
 The lattice size is $64^2$. }
\begin{center}
\begin{tabular}{||c|c|c|c|c|c|c||}
\hline
$\hspace{1em} \beta \hspace{1em}$  &
$\hspace{1em} \kappa \hspace{1em}$  &
$\hspace{1em} Q \hspace{1em}$  &
$\hspace{0.5em} N_p=0 \hspace{0.5em}$  &
$\hspace{0.5em} N_p=1 \hspace{0.5em}$  &
$\hspace{0.5em} N_p=2 \hspace{0.5em}$  &
$\hspace{0.5em} N_p=3 \hspace{0.5em}$  \\
\hline\hline
8.0  &  0.0  &  5  &  {\it 367.2}  &  {\it 372.2}  &  383.3  &
                                  \\
8.0  &  0.1  &  4  &  {\it 420.7}  &  {\it 403.5}  &  389.3  &
                      398.9       \\
8.0  &  0.2  &  7  &  {\it 353.3}  &  362.8        &  380.1  &
                                  \\
8.0  &  0.3  &  7  &  {\it 337.4}  &  355.1        &  368.3  &
                                  \\
8.0  &  0.4  & -3  &  {\it 316.3}  &  327.6        &  345.7  &
                                  \\
8.0  &  0.5  & -4  &  337.4        &  339.6        &  351.3  &
                                  \\
8.0  &  0.6  &  0  &  {\it 294.7}  &  {\it 278.0}  &  275.3  &
                      299.0       \\
8.0  &  0.7  & -2  &  {\it 263.2}  &  280.3        &  305.3  &
                                  \\
8.0  &  0.8  & -1  &  {\it 241.0}  &  262.2        &  285.4  &
                                  \\
\hline
\end{tabular}
\end{center}
\end{table}
 The uniquely determined minima are given in {\it italic}.
 The roman entries indicate that there are several local minima with
 close values or the minimum is not easy to find.
 Three additional pairs typically give substantially higher
 minima, which are usually not interesting and were therefore not
 precisely determined.

 In general, the question has to be answered what is to do if
 there are two or several almost equally (or even exactly) deep
 minima.
 A correct prescription for gauge fixing in this case is to take the
 average over these degenerate absolute minima.

 At the stronger gauge coupling $\beta=2$ with a plaquette
 expectation value $\simeq 0.7$ there are about 5-10 additional
 kink-antikink pairs at the best found minimum.
 The positions of these many kinks and antikinks are usually not
 sharply defined, there are many local minima with very similar values
 of $f_2$.
 Nevertheless the algorithm still works fine, the obtained minima are
 about a factor 3 smaller after the third step than after a standard
 gauge cooling in the first step.
 Due to the large number of kinks, however, the numerical work is
 increased substantially compared to $\beta=8$.


\section{Generalizations}                                 \label{s4}
 As the numerical experience shows, the gauge fixing algorithm based
 on the knowledge of the vacuum Gribov copies works well as long as
 the gauge coupling is not very strong.
 The decent range can be characterized by the plaquette expectation
 value being above $\simeq 0.6$, where the topological charge of
 the gauge configuration is reasonably well defined \cite{PROCS}.

 The generalization of this algorithm to higher (three or four)
 dimensional U(1) gauge theories is possible.
 Step 1. and 3. remain essentially the same.
 In step 2. one can remove the superfluous kink-antikink pairs
 first in parallel two dimensional subspaces and then remove the
 remaining closed strings of kinks between these ''slices'', first in
 three and then in four dimensions.

 The close analogy between the topological features in two dimensional
 U(1) and four dimensional SU(2) gauge theories is very helpful for
 a generalization to four dimensional SU(2).
 The r\^ole of the gauge field kinks in the vacuum is played there
 by vacuum gauge configurations on $2^4$ hypercubes which contribute
 by a non-zero integer to the representation of the topological charge
 in terms of the ''sections'' \cite{LUSCH,GKLSW}.
 Nevertheless, the numerical task seems considerably more difficult
 because of the non-linearity of the problem.
 I hope to return to this in a subsequent publication.

\hspace{3em}
%

\newpage

\begin{figure}
\vspace{9cm}
\includegraphics{ideal.kink.ps}
\caption{ \label{fig1}
 An idealized kink: the normalized link angles $t_{x\mu}$ are given.
}
\end{figure}

\begin{figure}
\vspace{11cm}
\includegraphics{ideal.kinkpair.ps}
\caption{ \label{fig2}
 An example of an idealized kink-antikink pair: kink on plaquette $k$
 and antikink on plaquette $a$.
 The normalized gauge transformation angles $a_x \equiv \alpha_x/\pi$
 are given, which create this pair from the identically zero vacuum.
 The ''overtransformed'' links are marked by double lines.
}
\end{figure}


\begin{thebibliography}{99}
%
\bibitem{WILSON}
K.G. Wilson,
Phys.\ Rev.\ \underline{D10} (1974) 2445.
%
\bibitem{GRIBOV}
V.N. Gribov,
Nucl.\ Phys.\ \underline{B139} (1978) 1.
%
\bibitem{ZWANZI}
D. Zwanziger,
Nucl.\ Phys.\ \underline{B209} (1982) 336;
Nucl.\ Phys.\ \underline{B378} (1992) 525.
%
\bibitem{VANBAA}
P. van Baal,
Nucl.\ Phys.\ \underline{B369} (1992) 259.
%
\bibitem{SALMHO}
M. Salmhofer,
Nucl.\ Phys.\ B (Proc.\ Suppl.) \underline{30} (1993) 81.
%
\bibitem{HULSEB}
A. Hulsebos,
Nucl.\ Phys.\ B (Proc.\ Suppl.) \underline{30} (1993) 539.
%
\bibitem{BOMIMUPA}
V.G. Bornyakov, V.K. Mitrjushkin, M. M\"uller-Preussker, F. Pahl,
Phys.\ Lett.\ \underline{B317} (1993) 596.
%
\bibitem{POYEZU}
M.I. Polikarpov, K. Yee, M.A. Zubkov,
preprint LSU-431-93.
%
\bibitem{DEGTOU}
T.A. DeGrand, D. Toussaint,
Phys. Rev. \underline{D22} (1980) 2478.
%
\bibitem{PROCS}
I. Montvay,
DESY preprints 93-134, 93-164 (1993); to appear in
{\it Proc. of the 17th Johns Hopkins Workshop}, Budapest, July 1993;
{\it Proc. of the 2nd IMACS Conference}, St. Louis, October, 1993;
{\it Proc. of the Lattice '93 Conference}, Dallas, October, 1993.
%
\bibitem{GKSW}
M. G\"ockeler, A.S. Kronfeld, G. Schierholz, U.-J. Wiese,
Nucl. Phys. \underline{B404} (1993) 839.
%
\bibitem{LUSCH}
M. L\"uscher,
Comm. Math. Phys. \underline{85} (1982) 39.
%
\bibitem{GKLSW}
A.S. Kronfeld, M.L. Laursen, G. Schierholz, U.-J. Wiese,
Nucl. Phys. \underline{B292} (1987) 330;     \\
M. G\"ockeler, A.S. Kronfeld, M.L. Laursen, G. Schierholz,
U.-J. Wiese,
Nucl. Phys. \underline{B292} (1987) 349.
%
\end{thebibliography}
\end{document}